\documentclass{ws-procs961x669}            
\usepackage{relsize}
\usepackage{mathrsfs}
\usepackage{multirow}
\usepackage{textcomp}
\usepackage{textcase}
\usepackage{siunitx}
\usepackage{amsmath}
\usepackage{hyperref}
\usepackage{amssymb}
\usepackage{footnote}

\newcommand{\dd}{\mathnormal{d}}
\newcommand{\old}[1]{}

\begin{document}

\title{Properties of accretion disc models in the background with quadrupoles}

\author{Shokoufe Faraji}

\address{%
 University of Bremen, Center of Applied Space Technology and Microgravity (ZARM), 28359 Germany
}%

\begin{abstract}
We consider a static and axially symmetric metric containing two quadrupole parameters. In the present contribution, we study the quadrupole moments constraints on the properties of the relativistic accretion disc models, also explore the relation of oscillatory frequencies of charged particles to the frequencies of the twin high-frequency quasi-periodic oscillations observed in some microquasars. We also compare the results with Schwarzschild and Kerr metrics.
\end{abstract}


\bodymatter

\section{Introduction}

In this paper, we are interested in studying how the presence of quadrupole moments in the metric affects the nature of astrophysical properties. This metric may link the observable effects to the system due to taking these parameters as the new dynamical degrees of freedom.

The first static and axially symmetric solution of Einstein equation containing arbitrary quadrupole moment were described in \cite{doi:10.1002/andp.19173591804}. Later, Zipoy and Voorhees \cite{doi:10.1063/1.1705005,PhysRevD.2.2119} found an equivalent transformation which can be treated analytically. Later on, with introducing a new parameter is known as $\rm q$-metric \cite{2011IJMPD..20.1779Q}. In this regard, quadrupoles can be seen as perturbation parameters of the Schwarzschild spacetime from a dynamical point of view. This metric represents the exterior gravitational field of an isolated static axisymmetric mass distribution. Of course, there are different ways of including quadrupole in the metric, but we consider ones that can be treated analytically. One simple and straightforward extension of q-metric involves introducing an external field in addition to the dynamical metric. This class of theories is referred to as generalized q-metric. The astrophysical motivation for choosing such fields is the possibility to constitute a reasonable model for a more realistic situation occurring in vicinity of compact objects.

This work studies thin and magnetized thick accretion disc models \cite{2013LRR....16....1A}, and quasi-periodic oscillations (QPOs) \cite{2019NewAR..8501524I} in this static and axisymmetric background.



\section{q-metric with an external source}\label{sp}

Most of the astrophysical studies considered the Kerr black hole. However, here, we choose to work on the generalized $\rm q$-metric that considers the existence of a static and axially symmetric external distribution of matter in its vicinity up to quadrupole like adding a magnetic surrounding \cite{1976JMP....17...54E}. In fact, due to their strong gravitational field, compact objects are not necessarily isolated or spherically symmetric as a response to their environments. This metric describes the outer of a deformed compact object up to quadrupole and reads as \cite{universe8030195}

\begin{align}\label{EImetric}
	{\rm d}s^2 &= - \left( \frac{x-1}{x+1} \right)^{(1+{\alpha})} e^{2\hat{\psi}} \dd t^2+ M^2(x^2-1) e^{-2\hat{\psi}} \nonumber\\
	 &\left( \frac{x+1}{x-1} \right)^{(1+{\alpha})}\left[ \left(\frac{x^2-1}{x^2-y^2}\right)^{{\alpha}(2+{\alpha})}e^{2\hat{\gamma}}\right. \nonumber\\
	 &\left. \left( \frac{\dd x^2}{x^2-1}+\frac{\dd y^2}{1-y^2} \right)+(1-y^2) \dd{\phi}^2\right],\
\end{align}
where $t \in (-\infty, +\infty)$, $x \in (1, +\infty)$, $y \in [-1,1]$, $\phi \in [0, 2\pi)$, and in principle $\alpha\in(-1,\infty)$. Metric function $\hat{\psi}$ plays the role of gravitational potential, and once we have it by an integration of its explicit form, $\hat{\gamma}$ is obtained. Up to quadrupole these functions are given by

\begin{align} \label{1111}
\hat{\psi} & = -\frac{\beta}{2}\left[-3x^2y^2+x^2+y^2-1\right],\\
\hat{\gamma} & = -2x(1+\alpha)\beta(1-y^2)\nonumber\\
  &+\frac{\beta^2}{4}(x^2-1)(1-y^2)(-9x^2y^2+x^2+y^2-1).\
\end{align}
By its construction, this metric is valid locally \cite{1982JMP....23..680G}. This metric has two parameters aside from the mass; deformation parameter $\alpha$ and distortion parameter $\beta$, which are not independent of each other. These are chosen to be relatively small and connected to the $\rm q$-metric and the surrounding external mass distribution, respectively. It can easily be checked that for vanishing $\beta$ we recover the $\rm q$-metric, and in the limits $\alpha=0$ and $\beta\neq 0$ distorted Schwarzschild written in the prolate spheroidal coordinates \cite{1982JMP....23..680G}, and for $\alpha=0$ and $\beta=0$ we recover the Schwarzschild solution. The relation between the prolate spheroidal coordinates $(t, x, y, \phi)$, and the Schwarzschild coordinates $(t, r, \theta, \phi)$ is

\begin{align}\label{transf1}
 x =\frac{r}{M}-1 \,, \quad  y= \cos\theta.\,
\end{align}
To the end, we explore this background by analyzing the thick and thin accretion disc models, and Quasi-periodic oscillations.

\section{Accretion discs}\label{ac}

In this analysis, we consider two well known Thin and Thick analytical models of accretion onto a compact object. 

\subsection{Thin model}

The standard thin disc model has been used to explain a variety of observations where gas is cold and neutral. Therefore, the coupling between the magnetic field and gas is negligible. In general, observations provide the luminosity and the maximum temperature of the disc that fit the data model.

In this model, the steady axisymmetric fluid configuration is assumed. In addition, as a result of the geometrically thin assumption, effectively, all physical quantities only depend on the vertical distance from the equatorial plane and the radial distance to the central object so that we can consider vertically integrated quantities. Therefore, the two-dimensional disc structure can be decoupled to one-dimensional configurations. One is responsible for the radial quasi-Keplerian flow another for the vertical hydrostatic structure \cite{1973blho.conf..343N,1974ApJ...191..499P}.

Three fundamental equations govern the radial structure of the thin disc model

\begin{align}\label{restmasscon}
&(\rho u^{\mu})_{;\mu}=0, \\
&u_{\mu} T^{\mu \nu}{}_{;\nu}=0,\\
&h_{\mu \sigma}(T^{\sigma \nu})_{;\nu}=0. \
\end{align}
where $u^{\mu}$ is the four-velocity of the fluid and $\rho$ is the rest mass density, $T^{\sigma \nu}$ is the stress-energy tensor containing non-vanishing shear part, and $h^{\mu \nu}$ is the projection tensor. These equations, together with the radial velocity, which is stated in terms of surface density and mass accretion rate, alpha viscosity prescription and heat flow, govern the radial structure of the disc \cite{1973blho.conf..343N,1974ApJ...191..499P}.

For the vertical profile in the comoving fluid frame, the force due to vertical pressure gradient is balanced with gravity, the centrifugal force, and vertically Euler force. The pressure is the sum of gas pressure from nuclei and the radiation pressure, which in practical is derived from the relativistic Euler equation \cite{1997ApJ...479..179A}.

\begin{figure}
\centering
\includegraphics[width=5.5cm]{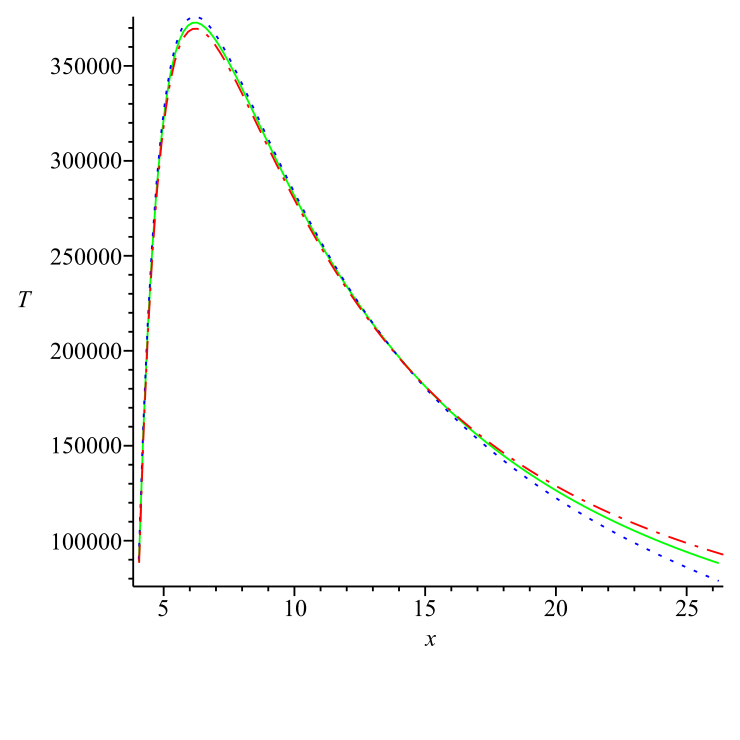}
\includegraphics[width=6cm]{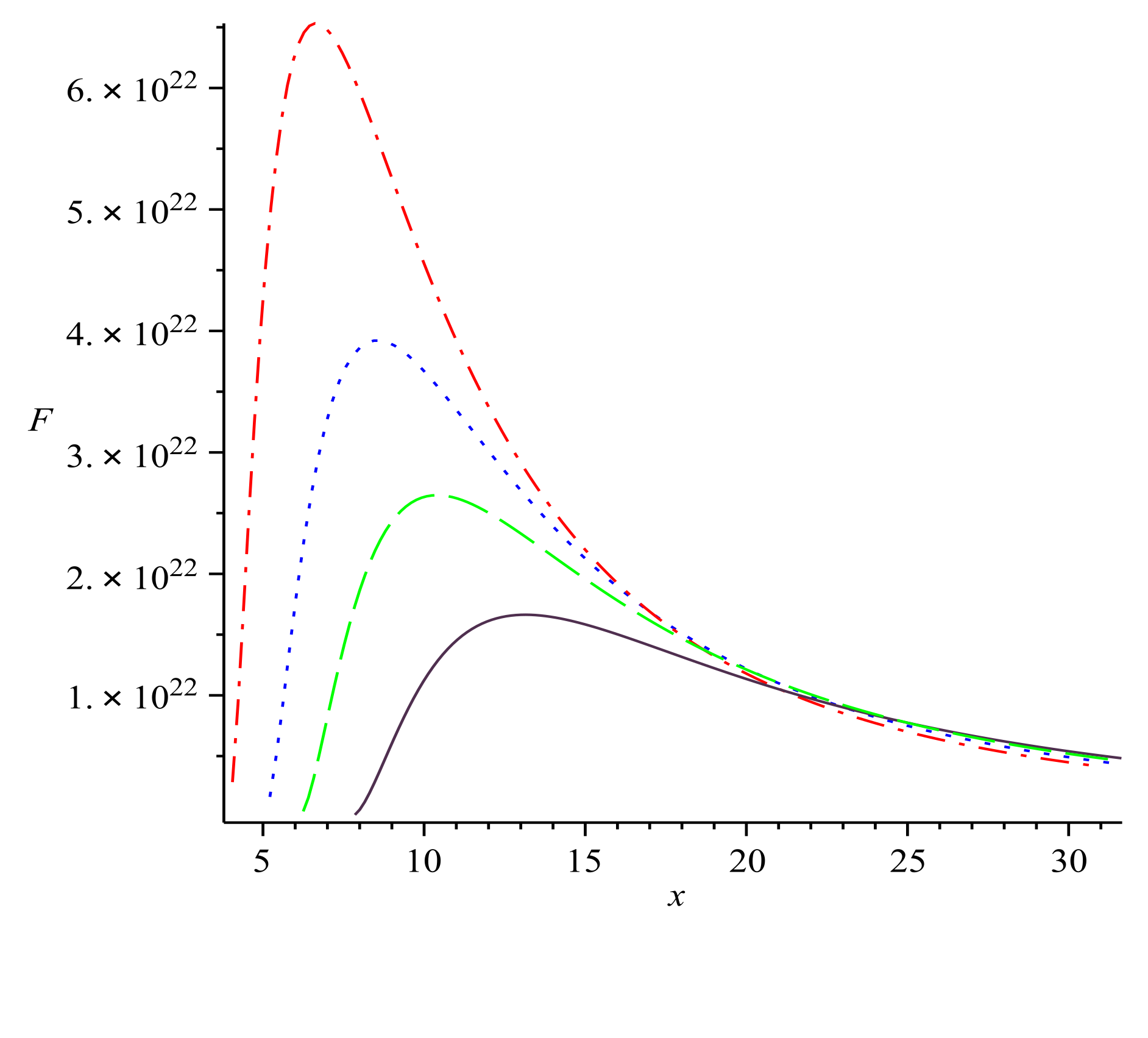}
\caption{\label{T}The left plot is temperaturee $T$ profile for $\alpha=-0.2$; blue line shows $\beta=-0.00001$, green one $\beta=0$ (Schwarzschild), and red one $\beta=-0.00001$. The right plot is flux profile $F$ for vanishing $\beta$ (q-metric) and different values of $\alpha$; from top $\alpha=-0.2$ , $\alpha=0$ (Schwarzschild), $\alpha=0.2$, $\alpha=0.5$.}
\end{figure}

Finally, by solving this system of linear equations, one obtains a unique solution for considering positive temperature and pressure. Besides quadrupoles, three parameters describe a thin disc solution in this background, namely $M$, mass accretion rate $\dot{M}$, and viscosity parameter $\alpha$. Consequently, the geometric configuration of an accretion disc located around this space-time depends explicitly on the value of the quadrupole parameter. An analysis shows the astrophysical quantities possess almost the same pattern. Figure \ref{T} shows the temperature and flux profiles for different values of quadrupole parameters. As it shows, the intensity of these quantities is much higher closer to the central object and for negative values than Schwarzschild and positive values. This links to the location of ISCO, which is considered the inner edge of the thin accretion disc. For negative quadrupoles, ISCO is closer to the central object, pushing away for positive ones. Therefore, for negative values, the inner part of the disc is located closer to the central object, which causes more intense characteristics in all quantities. It means the behaviour of positive quadrupole moments is smoother than negative ones in general. We may compare this result with the Kerr solution if we consider the rotation parameter a kind of quadrupole counterpart in this background. In the Kerr background, the intensity of these quantities for co-rotating is also smaller than the counter-rotating case. However, the main difference in the Kerr background is this pattern is valid everywhere, not only closer to the inner part of the disc. Therefore, in general, there are distinguishable differences in both cases regarding observational data   \cite{2020arXiv201106634F}.

\subsection{Thick model}

The thick disc model describes a general method of constructing perfect fluid equilibria in an axially symmetric and stationary background, which is the simplest analytical model of discs with no accretion flow and is radiatively inefficient. In this work, we considered its magnetic version. The evolution of an ideal magnetised fluid is described by the baryon conservation, energy-momentum conservation, and induction equation \citep{1989rfmw.book.....A,1978srfm.book.....D}. The final well known equation in this model reads as \cite{Komissarov_2006}  

\begin{align}\label{eq:FinalEq}
W-W_{\mathrm{in}}+\frac{\kappa}{\kappa-1}\frac{p}{w}+\frac{\eta}{\eta-1} \frac{p_m}{w}=\int_{\ell_{in}}^{\ell}{\frac{\Omega{\rm d}\ell}{1-\Omega\ell}},
\end{align}
where $W=\ln|u_t|$, magnetic pressure and gas pressure are $p_m$ and $p$, $\eta$ and $\kappa$ are parameters, and  $w$ is enthalpy. Therefore, one can construct this model by specifying $\Omega$ or $\ell$ to fix the geometry of the equipotential surfaces. In the constant case, the disc surface is fully determined by the choice of the potential in the inner part of the disc ,$W_{\rm in}$, independent of the magnetic field \cite{1989PASJ...41..133O}. Thus, the value of angular momentum ,$\ell_0$, determines the total potential \cite{Komissarov_2006}.

In this work we also considered two more angular momentum distributions; the power-law \cite{2015MNRAS.447.3593W}, and the trigonometric function \cite{2009A&A...498..471Q}. In the power-law distribution it is assumed that

\begin{align}
\Omega(\ell)=c\ell^{n}.    
\end{align}
where by using this equation, the right hand side of the equation \eqref{eq:FinalEq} can be calculated easily \citep{2015MNRAS.447.3593W}. In the trigonometric function distribution, it is assumed that \citep{2009A&A...498..471Q}

\begin{align}\label{triequ}
 \ell(x,y)=
   \left\{
  \begin{array}{@{}ll@{}}
  \ell_0\left(\frac{\ell_K(x)}{\ell_0}\right)^{\sigma}(1-y^2)^{\delta}, & x\geq x_{\rm ms}, \\
     \ell_0(\zeta)^{-\sigma}(1-y^2)^{\delta}, & x<x_{\rm ms},
    \end{array}\right.
\end{align}
where $\ell_0=\zeta \ell_K(x_{\rm ms})$, and $\ell_K$ is the Keplerian angular momentum in the equatorial plane, $\eta$, $\sigma$, and $\delta$ are parameters, and $x_{\rm ms}$ is the place of marginally stable orbit. In this model, one only needs to solve function $F$ to obtain the solution \cite{2009A&A...498..471Q}

\begin{align}
    \frac{\partial_xp}{\partial_{y}p}=\frac{\partial_xg^{tt}+\ell^2\partial_xg^{\phi\phi}}{\partial_{y}g^{tt}+\ell^2\partial_{y}g^{\phi\phi}}:= -F(x,y).
\end{align}
Analysis of results shows that changing parameter $\beta_c$ has a strong effect on the location and amplitude of the rest-mass density and spreading the matter through the disc for any chosen value of quadrupoles.

\begin{figure}
\centering
\includegraphics[width=6cm]{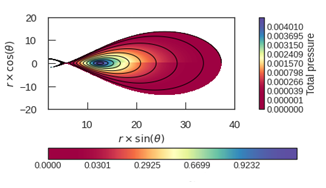}
\includegraphics[width=5.5cm]{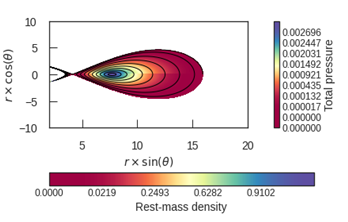}
\caption{The left plot is for $(\alpha,\beta)=(0.4,0.000001)$, the right one is for $(\alpha,\beta)=(-0.1,0.000001)$ with constant angular momentum.}
\end{figure}

\begin{figure}
\centering
\includegraphics[width=7cm]{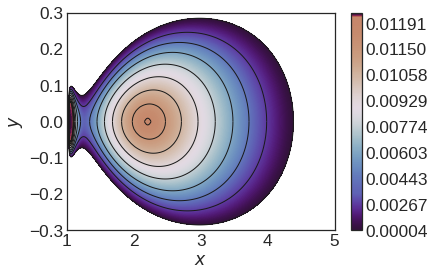}
\caption{\label{22}Possibility to have two tori in a row for $\alpha\in(-0.5,-0.553]$ regardless of the $\beta$ values.}
\end{figure}

In fact, a higher magnetic pressure causes the matter to concentrate more in the inner part of the disc. The rest-mass density maximum and its location also depend on the quadrupoles values. Although the effect of quadrupoles correlates with the other parameters, we can conclude that quadrupoles are more responsible for shifting the position of the disc and have a positive contribution to its radial extension. However, this effect for $\beta$ is not easy to see, as the values of quadrupole parameters that we chose in our models are minimal \cite{2021A&A...654A.100F,2021PhRvD.104h3006F}. Moreover, there is a possibility of having two tori in a row for any value of $\beta$ but negative values for the parameter $\alpha\in(-0.5,-0.553]$. In Figure \ref{22}  this situation is plotted with constant angular momentum. However, with two other angular momentum distributions discussed in the paper it is also possible to have two tori \cite{2021A&A...654A.100F}.

Furthermore, in the trigonometric distribution, the $\sigma$ and $\delta$ parameters influence each other. However, one can say that $\sigma$ is more responsible for the location and amplitude of the rest-mass density, and $\delta$ affects the distribution of equidensity surfaces. In the case of the power-law distribution, steeper angular momentum tends to shrink the disc and decrease the amplitude of the rest-mass density and pushes away the location of its maximum simultaneously.



\section{Quasi-periodic oscillations}\label{qpo}

Quasi-periodic oscillations (QPOs) of X-ray power spectral density have been observed at low (Hz) and high (kHz) frequencies \cite{2000ARA&A..38..717V,2006csxs.book..157M}. There have been many models to explain QPOs in the past years. Among them is the Relativistic Precession Model (RPM), which relates the twin-peak QPOs to the Keplerian and periastron precession frequency on an orbit located in the inner part of the accretion disc \citep{1998ApJ...492L..59S}. Regarding, the high frequencies QPOs ( the two-picks in Fourier power density spectra) are considered as the resonances between oscillation modes. Within the ratio $3:2$ epicyclic resonance, the resonant frequencies are identified with frequencies of radial and vertical epicyclic axisymmetric modes \cite{2001A&A...374L..19A,2001AcPPB..32.3605K,2005AN....326..864A}.

In this work, we are interested in the dynamics that occur when a static, axisymmetric central compact object is embedded in a weak uniform magnetic field of strength $B$ aligned with the central body's symmetry axis and described by the $\phi$-component of the vector potential \citep{PhysRevD.10.1680}

\begin{equation}\label{aphivector}
    A_{\phi}= \frac{1}{2}B (x^2-1) e^{-2\hat{\psi}}
	 \left( \frac{x+1}{x-1} \right)^{(1+{\alpha})}.
\end{equation}
In addition, the specific energy and angular momentum of the particle read as  \cite{2021arXiv210303229F}

\begin{align}
    &E = \left( \frac{x-1}{x+1} \right)^{(1+{\alpha})} e^{2\hat{\psi}} \frac{dt}{ds} ,\\
    &L  = (x^2-1) e^{-2\hat{\psi}}\left( \frac{x+1}{x-1} \right)^{(1+{\alpha})} \left(\frac{d \phi}{ds}+Q\right) ,\
\end{align}
where $Q:=\frac{qB}{2m}$ is magnetic parameter. Further, the effective potential is

\begin{align}\label{Vei}
    &V_{\rm Eff}=\left( \frac{x-1}{x+1} \right)^{(\alpha+1)} e^{2\hat{\psi}} \left[\epsilon+ \right.\\
    &\left.e^{-2\hat{\psi}} \left( \frac{x+1}{x-1} \right)^\alpha (1-y^2)
    \left(\frac{Le^{2\hat{\psi}}}{(x+1)(1-y^2)} \left( \frac{x-1}{x+1} \right)^\alpha - Q(x+1) \right)^2\right], \nonumber\,
\end{align}
The second term corresponds to the central force potential and electromagnetic potential energy. In what follows, we investigate the stability of circular motion in this background in the presence of a homogeneous magnetic field. We followed the method described in \cite{1981GReGr..13..899A,1986Ap&SS.124..137A} and extend the result to this background. The results are in good agreement with these papers regarding vanishing quadrupoles, also rotation in their works. 

We need to describe the more general class of orbits slightly deviated from the circular orbits in the equatorial plane $x^{\mu}$ using the perturbation expansion, $x^{\prime \mu}=x^{\mu}+\xi^{\mu}$. By considering this relation into the geodesic equation with non-vanishing external force and making the integration, for the $t$ and $\phi$ components, up to linear order in $\xi^{\mu}$, and we obtain \cite{1981GReGr..13..899A,1986Ap&SS.124..137A}

\begin{align}
   &\omega^2_{x} = \partial_x U^x - \gamma ^{x}_{\ \eta}\gamma ^{\eta}_{\ x} ,\\
   & \omega^2_{y} = \partial_y U^y .\
\end{align}
These equations describe radial and vertical oscillations of the charged particle about circular orbits. The sign of frequencies determines the dynamic $\omega^2_{x}$ and $\omega^2_{y}$ so that if they have a positive sign, we have a stable orbit; otherwise, even a minimal perturbation can cause a substantial deviation from the unperturbed orbit.
 
In Schwarzschild solution, the vertical epicyclic frequency is a monotonically decreasing function of $x$, and we have $\omega^2_x<\omega^2_y=\Omega^2$. On the contrary to the Schwarzschild case, the interesting situation in this background is to have various orderings among frequencies and thus different possibilities to reproduce the ratio of $3:2$, also other ratios via different combinations of parameters \cite{2021arXiv210303229F} which is not the case in either Schwarzschild or in $\rm q$-metric.

Among the various models of the resonance of accretion disc oscillations, we consider the group of QPO models (WD, TD, RP, Ep, Kp, RP1, RP2) considered in \citep{2011A&A...531A..59T} and examined them in this background and compared with observational data of three microquasars  XTE 1550-564, GRS 1915+105, and GRO 1655-40 \cite{2021arXiv210303229F,2021arXiv210211871F}. However, the physical details of these models are different and depend explicitly on the time evolution of the desired system. Indeed, the case of relatively slowly rotating XTE 1550-564 source is more compatible with our static set-up. However, different sets of parameters can fit the data even for the two other fast-rotating sources. An analysis shows for chosen parameters $Q$, $\alpha$, and $\beta$, the best fit almost corresponds to RP2, RP, and RP1 models in low spin cases. Besides, results reveal that even the regular orbits for some combinations of parameters of metric turn to behave chaotically with a higher magnetic field.

\begin{figure}
    \centering
    \includegraphics[width=0.45\hsize]{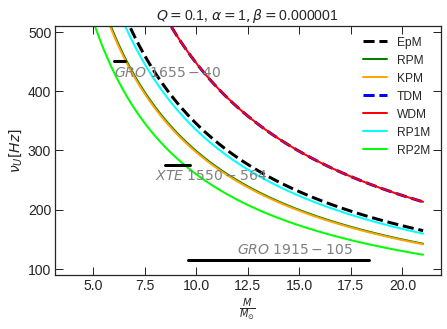}
    \includegraphics[width=0.45\hsize]{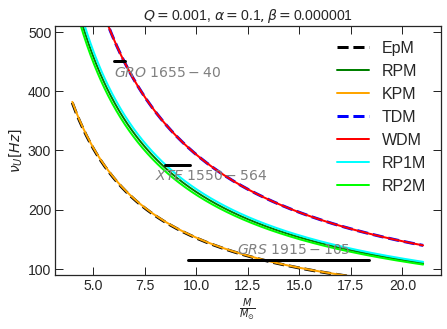}
    \caption{The upper oscillation frequencies compared to the mass limits obtained from observations of three microquasars for some chosen parameters.}
    \label{fig16Data}
\end{figure}

\section{Conclusion}

In this paper, we mentioned astrophysical features of two accretion disc models in the background of a compact object containing quadrupole moments. We have shown that the presence of quadrupoles changes the accretion disc's geometric properties in both models drastically. In particular, in the Thin accretion disc model, the spectral features would be significantly different and distinguishable in the observation from a black hole. The structure and the shape of the magnetized Thick accretion disc model are also influenced strongly by these parameters. 

In addition, the dependency of these two parameters reflects into epicyclic frequencies of particles that cost substantial deviation from the correspondence quantities in the Schwarzschild case. In fact, the resonant phenomena of the radial and vertical oscillations at their frequency ratio $3:2$ depending on chosen parameters, also in this background can be well related to the frequencies of the twin HF QPOs $3:2$ observed in the microquasars.  Although the fitting depends on the combination of parameters, we still can expect to have a better fitting for choosing a larger quadrupole moment in the combination.

Directions for the future could be adding the strong magnetic field which also influences the metric itself, considering rotation, studying the self-gravity of the Thin disc model, and studying the stability of this solution to extend this work. Further, this theoretical model can be served as the initial data for numerical simulations in the astrophysical setting.

\bibliographystyle{ws-procs961x669}
\bibliography{accrectionp}

\end{document}